# Generalizability of deep learning–based fluence map prediction as an inverse planning approach

Lin Ma[1], Mingli Chen[1], Xuejun Gu[†1] and Weiguo Lu[†1]

[1]*Medical Artificial Intelligence and Automation Laboratory, Department of Radiation Oncology, University of Texas Southwestern Medical Center, 2280 Inwood Rd, Dallas, TX 75390, USA*

**Abstract**

Deep learning–based fluence map prediction (DL-FMP) method has been reported in the literature, which generated fluence maps for desired dose by deep neural network (DNN)–based inverse mapping. We hypothesized that DL-FMP is similar to general fluence map optimization (FMO) because it's theoretically based on a general inverse mapping. We designed four experiments to validate the generalizability of DL-FMP to other types of plans apart from the training data, which contained only clinical head and neck (HN) full-arc volumetric modulated arc therapy (VMAT) plans. The first three experiments quantified the generalizability of DL-FMP to multiple anatomical sites, different delivery modalities, and various degree of modulation (DOM), respectively. The fourth experiment explored the generalizability and stability to infeasible dose inputs. Results of the first experiment manifested that DL-FMP can generalize to lung, liver, esophagus and prostate, with gamma passing rates (GPR) higher than 95% (2%/2mm). The second experiment showed that DL-FMP can generalize to partial-arc plans and predict partial-arc fluences. GPR (3mm/3%) ranged from 96% to 99%. DL-FMP cannot generate fluence maps in discrete beam angles for IMRT input. But the predicted dose still agreed with ground truth dose with 93% GPR (5%/5mm). The third experiment demonstrated that DL-FMP can generalize to various DOMs, with GPRs (3%/3mm) ranged in 94%-98%. Moreover, the DOM of predicted fluence maps correlated to the optimality of the input dose accordingly. The fourth experiment exemplified that DL-FMP can make stable predictions for infeasible dose input. In conclusion, we validated that DL-FMP can generalize to plans for multiple anatomical sites, plans of different delivery modalities and plans with various DOM. It can also make stable prediction for infeasible input. Supported by its theoretical basis, we experimentally manifested that DL-FMP is a general inverse planning approach that can be applied to various clinical scenarios.

**Symbols**

$d$ Vectorized dose distribution with $N$ voxels. $d_i$ is dose in voxel $i$. The space of dose vectors is a subspace of $N$ dimensional non-negative real-valued vector space. $\{d\} \subset R_{0+}^N$

$w$ Vectorized fluence maps with $M$ beamlets. $w_j$ is intensity of beamlet $j$. Vectorized fluence maps belong to $M$ dimensional non-negative real vector space. $w \in R_{0+}^M$

$D$ Dose influence matrix with $N$ rows and $M$ columns. Each entry $D_{ij}$ is the full dose delivered to voxel $i$ by beamlet $j$, including contributions of primary dose and scatter dose.

$\widetilde{D}$ Approximate dose influence matrix involving only primary dose. $\widetilde{D}_{ij} = 0$ if voxel $i$ is not on the ray-tracing line of beamlet $j$.

$P$ Vectorized projections of dose, calculated by weighted summation of dose along beamlet ray-tracing lines. $P = \widetilde{D}^T \times d$. $\{P\} \subset R_{0+}^M$

## 1 Introduction

Inverse planning is the central topic of treatment planning of modern radiation therapy (Brahme, 1988), in which process planners inversely obtain machine parameters from the desired dose distribution through optimization. The advent of new treatment machines and delivery modalities allows the accurate control of millions of delivery parameters, which can be obtained only by optimization. Fluence map optimization (FMO) (Shepard et al., 2000, Lu, 2010) is designed for optimizing weights of beamlets for intensity modulation radiation therapy (IMRT) and direct machine parameter optimization (DMPO) (Lu, 2010, Otto, 2008) is mainly adopted for volumetric modulated arc therapy (VMAT). These optimization-based inverse planning approaches, albeit focused on generating delivery parameters, actually determined both the delivery parameters and admissible dose simultaneously. A non-trivial task of inverse planning is to find an admissible dose as optimization objective, which was manually adjusted in an inefficient try-and-error fashion. To overcome this issue, planning tools such as knowledge-based planning (KBP) and multi-criteria optimization (MCO) have been developed and integrated into treatment planning system (TPS) for an efficient operation of plan optimization. KBP (Appenzoller et al., 2012, Yuan et al., 2012) assumed that dose volume histograms (DVHs) of admissible doses can be inferred from geometric distribution of planning target volumes (PTVs) and organs at risk (OARs). MCO (Craft et al., 2012) relied on optimization to construct a Pareto surface, providing more than one admissible doses to users for real-time trade-off exploration.

The magic learning ability of deep learning (DL), specifically deep neural network (DNN), has been investigated for the application in treatment planning. Researchers demonstrated that DNN can learn the mapping from three-dimensional organ contours to 3D dose distribution as an KBP approach in various sites, such as head and neck (HN) (Babier et al., 2020, Nguyen et al., 2019b), lung (Barragán‐Montero et al., 2019), esophagus (Barragán-Montero et al., 2021, Zhang et al., 2020), pancreas (Wang et al., 2020), prostate (Nguyen et al., 2019c). More recently, DNN has been employed to generate Pareto-optimal dose distributions from patient geometry along with inputs for trade-off selection (Nguyen et al., 2019a, Ma et al., 2019, Nilsson et al., 2021). These DL models achieved high accuracy in voxel level and dosimetric parameters in regions of interest (ROIs). They can provide an admissible volumetric dose to TPS as optimization objective, but they cannot replace inverse planning since other half of job, obtaining delivery parameters, is still fulfilled by optimization. Voxel-based dose mimicking optimization or DVH-based optimization were adopted to convert the DL-predicted dose to delivery parameters (McIntosh et al., 2017, Babier et al., 2020, Nilsson et al., 2021). However, relying on optimization to restore delivery parameters fails to fully utilize the complete information in spatial dose distribution and prolongs planning time. Using FMO as a case, the dose $d$ is the product of dose influence matrix $D$ and fluences $w$, that is $d = D \times w$. Ideally, given spatial dose distribution $d$, the corresponding fluences $w$ can be derived by applying inverse mapping of $D$ to the given $d$ without time-consuming optimization. Developing a method that maps dose to fluence can fulfill the other half of job of FMO. Ma et al. (2020) proposed a DL-based inverse mapping method to predict fluence maps for desired VMAT dose distributions, which was supported by a general theoretical basis. This fluence map prediction method (DL-FMP) was trained and tested by clinical HN full-arc VMAT plans. What's more, the generalizability to prostate plans was also demonstrated (Ma et al., 2020).

Generalizability issue of DL models has been widely studied in medical physics. Imprudent application of an external / un-commissioned DL model to clinical use can cause severe adverse clinical consequences because of unknown failure modes and overlooked generalizability issue. We review the application and generalizability issue of DL models in four exemplary tasks in the field of medical physics below. The first task, disease diagnosis, is to identify disease patterns from radiological images and to relate the identified patterns to final diagnosis. Image classification DL models have been adopted for disease diagnosis because it functions in a similar identification-classification logic, as revealed in the work of "class activation map" (Zhou et al., 2016). For example, CheXNet (Rajpurkar et al., 2017) detected the abnormal lung region and subsequently gave diagnosis for 14 types of disease with radiologist-level accuracy. Although DL models can achieve human-performance in a confined dataset, the generalizability to new dataset is not-guaranteed. Nguyen et al. (2021) studied the generalizability of DL model for COVID-19 classification using four CT datasets from different countries. The key finding is that the model trained by a single dataset, though performed well in the same dataset, can only give random guess for the CT images from other datasets. They concluded that the bad generalizability was likely to be resulted from factors about data heterogeneity, including but not limited to patient demographics and differences in image acquisition or reconstruction. The second task, organ segmentation, is to localize organ of interest with the knowledge of human anatomy and to delineate organ boundary by image contrast. The famous image segmentation model, U-Net (Ronneberger et al., 2015), combined the deep-level semantic information and shallow-level boundary information to segment biomedical images. Further developments on model architectures considered the positional hierarchy of organs in HN (Chen et al., 2019) and prostate (Balagopal et al., 2018), which achieved good segmentation accuracy in the testing data from the same source. However, Feng et al. (2020) found that the organ (thoracic OARs) segmentation model trained by one dataset (2017 AAPM Thoracic Auto-segmentation Challenge) cannot generalize to their local dataset. They ascribed the failure of generalization to the systematic differences between the datasets: the CT images in the training datasets were acquired with free-breathing technique while the images in their local dataset used abdominal compression technique. The subtle shift of thoracic organs due to the abdominal compression caused significantly worse performance on the local dataset. The third task, CBCT-to-CT conversion, is a voxel-by-voxel regression that removes artifact and improves HU accuracy for CBCT images. Liang et al. (2019) implemented an image conversion DL model based on Cycle-GAN (Zhu et al., 2017) for this task and achieved results better than traditional registration method. They furthered investigated the generalizability of DL model and noticed that the performance degraded as the model was applied to images from different anatomical sites or from different scanners (Liang et al., 2020). The fourth task, dose prediction, is also a voxel-by-voxel regression that restores spatial dose distribution from the geometrical positions of ROIs. Kandalan et al. (2020) studied dose prediction DL models for prostate VMAT treatment and found that model performance degraded as it was evaluated by plans with planning styles different from training data. In all the tasks above, DL models are employed to learn the implicit mapping from input to output, which varies with data distribution. Using dose prediction as an example, we have to customize DL models for each anatomical site, planning styles (trade-off selection) and delivery modality (IMRT/VMAT) because all of them affect the mapping from ROI distribution to dose distribution.

This paper focuses on the generalizability of DL model for fluence map prediction and specially, the DL-FMP proposed in (Ma et al., 2020). We will demonstrate the generalizability of DL-FMP by theoretical analysis and then validate it by four experiments. The rest of paper is organized as follows. In the Methods and Materials section, we describe the principles of DL-FMP and experimental validation on its generalizability. Section 2.1 articulates the theoretical basis of DL-FMP and justifies its generalizability in theory. Section 2.2 introduces the overall workflow and the design of experiments. Sections 2.3 details the four experiments separately: generalizability to multiple anatomical sites (2.3.1), generalizability to different delivery modalities (2.3.2), generalizability to plans with various degree of modulation (DOM) (2.3.3), and generalizability to infeasible dose objectives (2.3.4). In the Results section, we report the prediction accuracy of DL-FMP in the four experiments and quantify the generalizability. In the Discussion section, we further investigate the experiment results and conclude that DL-FMP can generalize to various clinical scenario for broad implementation.

## 2 Materials & Methods

### 2.1 Theory

The theory of fluence map prediction (Ma et al., 2020) is recapitulated in this section. The central topic of deep learning-based fluence map prediction is to learn the inverse mapping (1) from dose to fluence $f_d: d \mapsto w$ by a DNN.

$$f_d: d = Dw \mapsto w, \quad s.t. f_d \circ D = I \tag{1}$$

$I$ stands for identity mapping. DNN has been employed in lots of tasks where a local image-to-image mapping could be established, such as image conversion and image segmentation. However, $f_d$ is a highly non-local mapping. The intensity change in one beamlet of $w$ can alter the doses in thousands of voxels around the beamlet direction across the whole patient volume. Therefore, $f_d$, as a global mapping, is hardly learnable to a DNN who has dose as input and fluence as output.

Then we investigate if the mapping (2) from projections of dose to fluences, $f_P: P \mapsto w$, is learnable to a DNN.

$$f_P: P = \widetilde{D}^T D w \mapsto w, \quad s.t. f_P \circ \widetilde{D}^T D = I \tag{2}$$

Projections of dose $P$ are defined as $P = \widetilde{D}^T d = \widetilde{D}^T D w$. $P$ and $w$ are both defined in the isocenter plane of beam eye view (BEV) coordinate system (Lu, 2010). Their correlation is local in that the intensity change of one beamlet only dominates the change of several dose projections in its vicinity. So $f_P$, as a local mapping, is learnable to a DNN with projections of dose $P$ as input and fluences $w$ as output. However, our initial attempts on training a DNN to learn $f_P$ by clinical plans failed. On reflections, we realized that the projections of dose $P_{pt}$ and fluences $w_{pt}$ in clinical plans are defined for patient geometry (electron density distribution of patient body).

$$f_P: P_{pt} = (\widetilde{D}^T D)_{pt} w_{pt} \mapsto w_{pt}, \quad s.t. f_P \circ \left(\widetilde{D}^T D\right)_{pt} = I \tag{3}$$

So the inverse mapping $f_P$ (in patient geometry) DNN tried to learn satisfies $f_P \circ \left(\widetilde{D}^T D\right)_{pt} = I$. Since dose influence matrix $D_{pt}$ varies with electron density distribution of each patient, the inverse mapping $f_P$ for patient geometry is patient-specific. When we directly used clinical plans for training, data pairs $(input: P_{pt},\ output: w_{pt})$ were collected from different patients. So the mappings $f_P$ to learn were not consistent and the training failed to converge. Even though we may generate enough data (plans) for a single patient and train a DNN for this patient, the trained DNN cannot generalize to other patients since the electron density distribution varies and so do $\left(\widetilde{D}^T D\right)_{pt}$ and $f_P$.

To remove the dependence of $f_P$ on patient-specific electron density distribution, we proposed to learn the mapping $f_P$ (4) in a consistent phantom geometry.

$$f_P\colon P_{ph} = (\widetilde{D}^T D)_{ph} w_{ph} \mapsto w_{ph}, \quad s.t. f_P \circ \left(\widetilde{D}^T D\right)_{ph} = I \tag{4}$$

Clinical plans in various patient geometries were converted to a consistent cylindrical water phantom placed on isocenter with 15cm radius (Ma et al., 2020). This is achieved by a plan scaling technique (Ma et al., 2020), which is illustrated by purple arrows in figure 1. The mapping $f_P$ in phantom geometry satisfies $f_P \circ (\widetilde{D}^T D)_{ph} = I$. Since $D_{ph}$ is determined by a fixed phantom electron density distribution, the training data pairs $(input\colon P_{ph},\ output\colon w_{ph})$ are consistent and the DNN can learn the $f_P$ in phantom geometry, which is inverse to $\left(\widetilde{D}^T D\right)_{ph}$. We should note that $D_{ph}$ is also independent of data distribution, so the $P_{ph}$ and $w_{ph}$ in plans for multiple sites, in plans of different delivery modalities and in plans with various DOMs are linked by the same mapping $f_P$. Therefore, a DNN that learned the $f_P$ in phantom geometry can generalize to various types (site, delivery modality and DOM) of plans, which are not confined to the plans in training dataset.

## 2.2 DL-FMP workflow and experimental design

The workflow of DL-FMP is presented in figure 1. It starts from dose in phantom geometry, $d_{ph}$, whose projections $P_{ph}$ are the input of DNN. The DNN learned the inverse mapping $f_P\colon P \mapsto w$ for phantom geometry and maps $P_{ph}$ to corresponding fluences in phantom geometry, $w_{ph}$. Finally, we use a fluence map scaling scheme (Ma et al., 2020), which adapts to the geometrical variation from phantom to patient, and obtain the fluence maps for patient geometry $w_{pt}$, which can deliver the input desired dose in patient geometry.

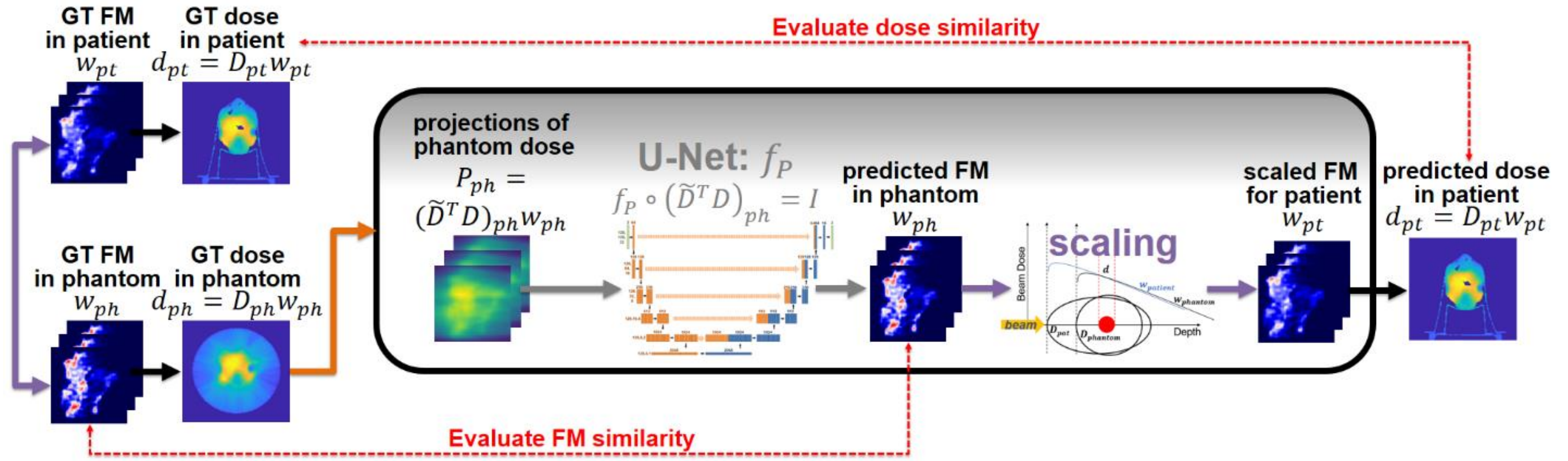

Figure 1. Overall workflow of fluence map prediction. Two steps of DL-FMP, DL model (U-Net) mapping and fluence map scaling, are in the black-shaded bounding box. Plan (dose and fluence map) in patient geometry (top left) is scaled to phantom geometry (bottom left) and input to DL-FMP. The prediction accuracy of DL-FMP is evaluated in terms of fluence (bottom red arrow) and dose (top red arrow) as compared to ground truth (GT).

We have demonstrated the theoretical basis for the generalizability of DL-FMP. Here we introduce the design of four experiments, which validate the generalizability of DL-FMP by testing plans apart from the plans in training dataset. The DL-FMP, specifically the DNN in the first step, was only trained by clinical

HN full-arc VMAT plans in (Ma et al., 2020). In section 2.3.1, we use testing plans for esophagus, liver and prostate (figure 2.a), which are anatomical sites different from training site (HN). In section 2.3.2, we use partial-arc VMAT plans and IMRT plans (figure 2.b), whose delivery modalities are different from that of full-arc training data. In section 2.3.3, we use plans with various DOM (figure 2.c). The training data are clinical plans with adequate DOM that can balance machine deliverability and dose optimality (Craft et al., 2007). The testing plans will cover a wide range of DOM, from one extreme that is too high to be deliverable, to another extreme which is so low that plan quality is not approvable. In section 2.3.4, we research how DL-FMP responds to an infeasible dose input. The question that whether DL-FMP will crash or still make stable fluence predictions will be answered. In all four experiments, the performance of DL-FMP will be evaluated by the prediction accuracy of fluence maps (in phantom geometry, bottom red arrow in figure 1) and doses (in patient geometry, top red arrow in figure 1). The fluence map prediction accuracy will be evaluated by similarity metrics customized to each experiment. The dose accuracy will be quantified by the gamma passing rate (GPR) between predicted doses (calculated from predicted fluence maps) and ground truth (GT) doses, and other metrics if applied.

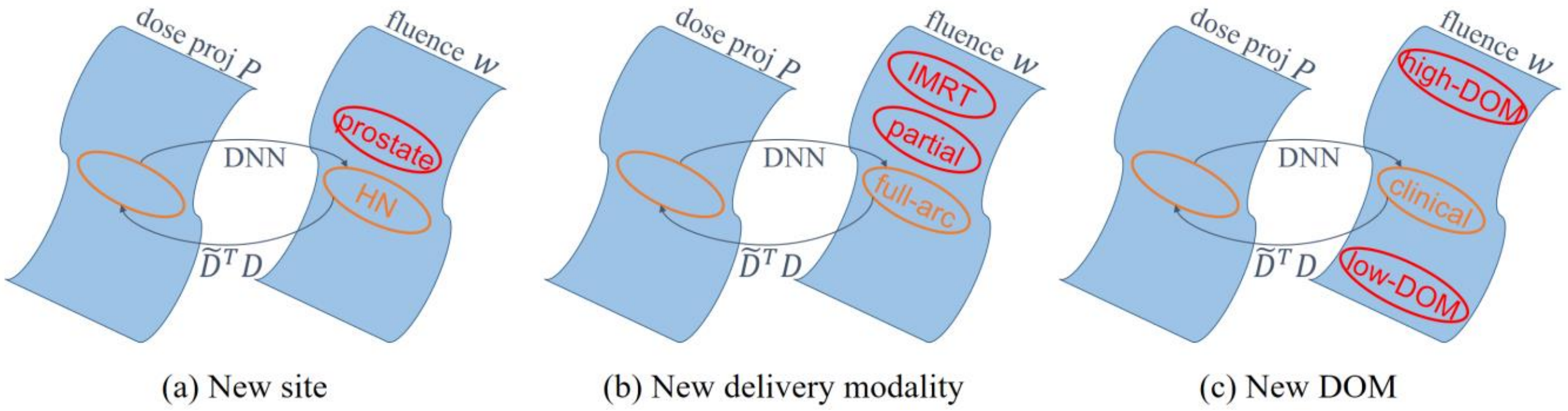

Figure 2. Schematic diagrams of experiments on the generalizability of DL-FMP. (a) Generalizability to multiple anatomical sites. (b) Generalizability to different delivery modalities. (c) Generalizability to various degree of modulations (DOMs). The open orange circle on the “dose proj P” manifold represents the data distribution of dose projections in training dataset. The experiment on infeasible dose is not shown in this figure.

## 2.3 Experimental data

### *2.3.1 Experiment on multiple sites*

To validate the generalizability of DL-FMP to multiple anatomical sites beyond the training site (HN), we evaluated the prediction accuracy of DL-FMP on four other sites: lung, liver, esophagus and prostate. All plans are 6MV coplanar full-arc VMAT plans, which differ from training plans only in site. The lung cancer plan (figure 3.a) has a PTV in the upper lobe of right lung, with a prescription dose of 60Gy and a volume of 466cc. The liver plan (figure 3.b) is a stereotactic body radiation therapy (SBRT) plan for a 50cc liver metastasis, which was prescribed 54Gy. The esophagus cancer plan (figure 3.c) is a simultaneously integrated boosted (SIB) plan with two PTVs. The boosted volume for primary tumor (50Gy, 196cc) is fully encompassed by another volume for micro-metastasis and lymph nodes (45Gy, 758 cc). The test on prostate cancer were made on 14 cases, which was reported in (Ma et al., 2020).

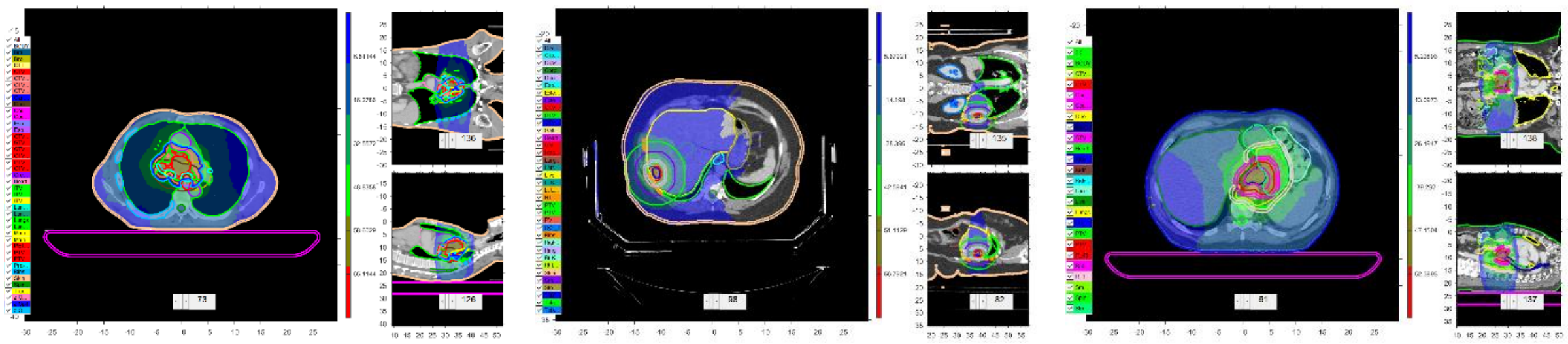

(a) Lung case. (b) Liver case. (c) Esophagus case.

Figure 3. Plans for multiple anatomical sites apart from the training site (HN). Each subfigure has contours and dose color wash overlaid on CT, with axial view (left), coronal view (top right) and sagittal view (bottom right).

### 2.3.2 *Experiment on different delivery modalities*

To validate the generalizability of DL-FMP to delivery modalities other than full-arc coplanar VMAT (the delivery modality of training data), we prepared three plans belonging to two new delivery modalities: partial-arc coplanar VMAT and coplanar IMRT. The first partial-arc plan is a clinical partial-arc VMAT plan for a HN case (figure 4.a). This plan has three arcs and the angular coverage of all arcs is from 212° to 350° (major arc) in total. PTV is at the left side of patient with a volume of 750cc. The second partial-arc plan is a clinical partial-arc VMAT plan in pelvis region (figure 4.b). This plan has two arcs and the angular coverage of all arcs is from 60° to 181° (major arc) in total. PTV is a single bone metastasis of prostate cancer at the right side of body with a volume of 16cc. These two partial-arc VMAT plans are more sparse and discontinuous than training cases (full-arc VMAT plans) in terms of the beam angles used by the plans. To test where the limit of DL-FMP is, we prepared an IMRT plan, which is of extreme angular sparsity and discontinuity compared to full-arc VMAT plan. We re-optimized an IMRT plan for a HN case (figure 4.c) using in-house NVBB-based optimizer. Prescription doses for three PTVs are 56Gy, 59.4Gy and 70Gy. The original design of DL-FMP used 64 uniformly distributed control point from 2.8º to 357.2º with a spacing of 5.6º (360º/64) to approximate the continuous arc delivery of VMAT plan (Ma et al., 2020). To make the test IMRT plan compatible with DL-FMP, we set the beam angles of IMRT plan as eight equally-spaced angles on the angular grid of DL-FMP. Although this beam configuration has unfavorable opposite beams, it serves for the purpose of testing DL-FMP with an angularly-sparse IMRT plan.

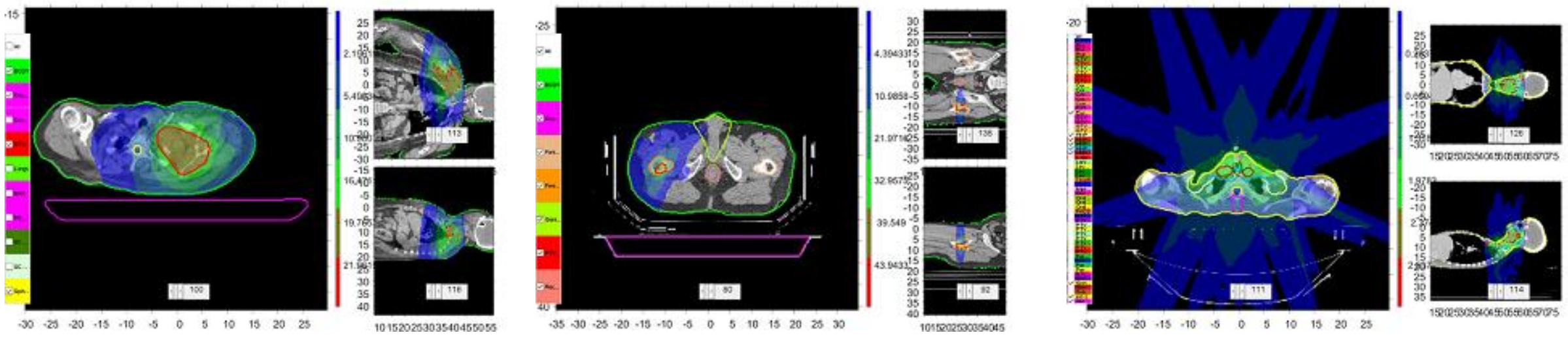

(a) HN partial-arc VMAT (b) Pelvis partial-arc VMAT (c) HN 8-beam IMRT

Figure 4. Experiment on different delivery modalities. Each subfigure has contours and dose color wash overlaid on CT, with axial view (left), coronal view (top right) and sagittal view (bottom right).

### 2.3.3 Experiment on various DOM

Plan quality (optimality of dose) can be improved by increasing DOM (Craft et al., 2007). The DOM of a plan reflects how hard we push the machine to its limit. Within reasonable treatment time, the DOM is bounded by machine constraints like maximum leaf speed. The trade-off between DOM and dose optimality in clinical plans balances deliverability and quality.

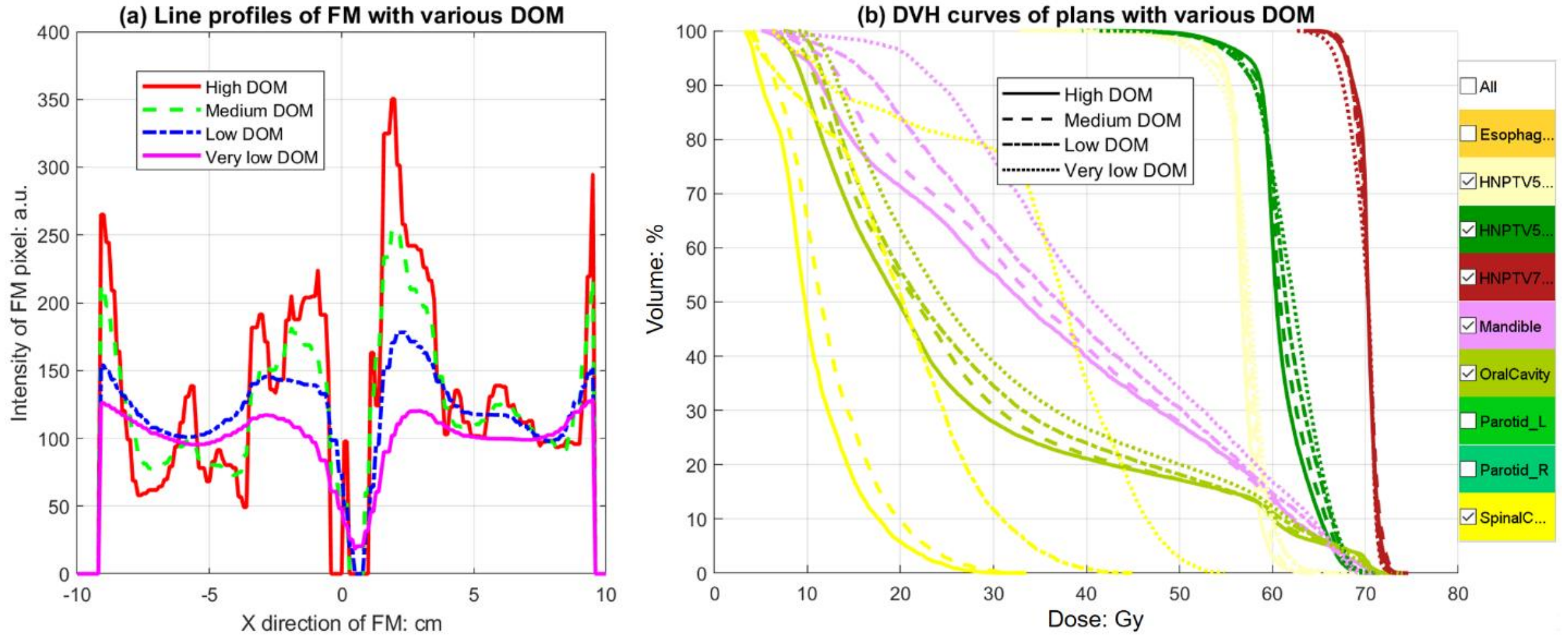

Figure 5. Experiment on various DOMs. (a) Comparison of intensity profiles created by one leaf pair in four plans with various DOMs. The central dent (around 0cm) is at the BEV position of spinal cord. (b) Comparison of DVH curves. Curves of some OARs (unchecked in ROI control panel) are not shown for a clear display.

The training data of the DNN inside DL-FMP are extracted from clinical VMAT plans, which reflected the clinically-desired trade-off between the DOM and dose optimality. To validate the generalizability of DL-FMP to plans with various DOMs, we re-optimized four plans with different DOMs for a HN case by NVBB optimizer. As a surrogate of actual VMAT plans, the four plans are 64-beam IMRT plans with beam angles on the angular grid of DL-FMP. Planning images and structures of this HN case are the same as the one in figure 4.c. There are three PTVs with prescription dose levels of 56Gy, 59.4Gy and 70Gy. Six organ at risks (OARs) are considered. The optimization objective was comprised of a dose objective and a regularization term on the DOM. We used smoothness of fluence map as the metric of DOM in plan optimization. Four plans were optimized with the same hyper-parameters, such as the objective dose and weighting of all nine ROIs, except for the weighting of regularization term on DOM. The weighting of DOM in optimization increased from plan #1 to plan #4. As a result, the plan #1, which received least regularization on DOM, had the highest DOM and the plan #4 had the lowest DOM. We refer to the four plans as high-DOM, medium-DOM, low-DOM and very low-DOM plan below. The profiles of fluence maps of the four plans are compared in figure 5.a and the DVHs are compared in figure 5.b. The ROI dose statistics is shown in table 4. As there is trade-off between DOM and plan quality, the plans with higher DOM achieved better OAR sparing and PTV homogeneity than plans with lower DOM. The profile and image of fluence map in high-DOM plan are shown in figure 5.a and figure 11 (top left), respectively, featuring by a highly-modulated pattern that is hard to be delivered. Meanwhile, the plan quality of very low–DOM plan is much worse than other three plans, reflected by table 4 and DVH curves in figure 5.b. So the high-DOM plan and the very low–DOM plan define a wide range of DOM, and represent the two extreme conditions that cannot co-exist in the training dataset of DL-FMP (clinically-approved plans).

### *2.3.4 Experiment on infeasible dose*

The DL-FMP was previously trained and tested with feasible dose distributions that were achieved by photon plans. In other words, the doses input to DL-FMP have corresponding photon fluences to achieve them. The optimality of photon dose distribution, measured by the PTV coverage and OAR sparing, is bounded by the physical properties of photon. Photon fluence has to be non-negative so a dose distribution with uniform dose in PTV and zero dose outside is not achievable for photon. Moreover, photon attenuates by negative exponential of penetration depth, so photon beam must deposit harmful dose to tissue behind tumor. As a result, proton dose distributions are not feasible for photons in that proton can stop at the distal margin of tumor but photons never stop completely. In the two examples of infeasible dose above, the dose falloff outside of target is too fast to be achieved by photon beam. The total dose inside phantom is too low to achieve the target dose so energy conservation law is violated.

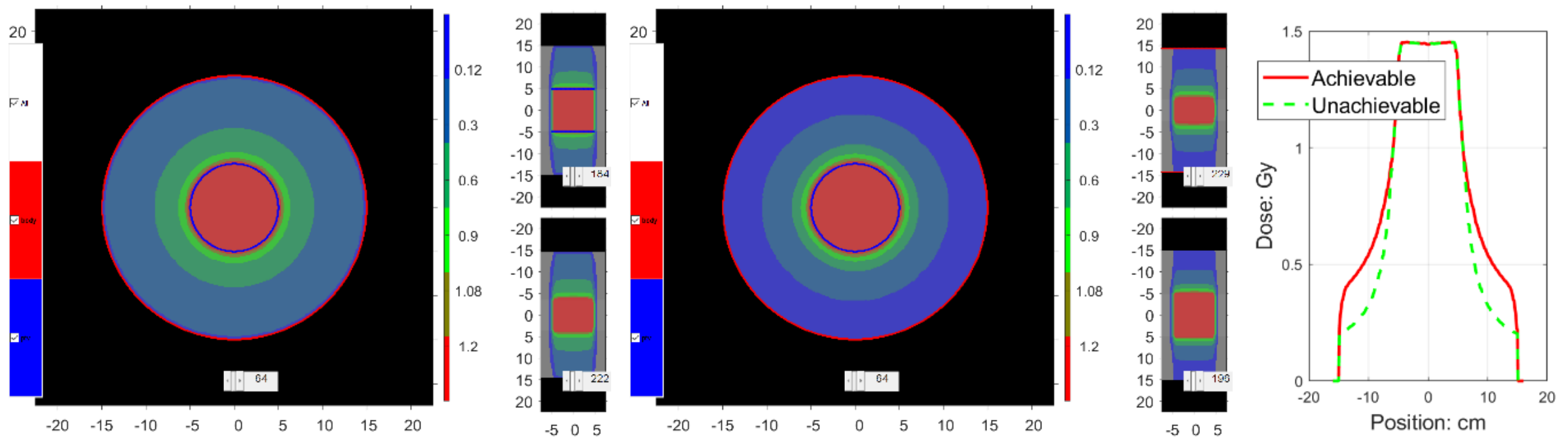

(a) Feasible dose (b) Infeasible dose (c) Profiles

Figure 6. Experiment on infeasible dose. (a) Feasible (achievable) dose. (b) Infeasible (unachievable) dose. Two doses are displayed with the same color scheme. (c) Comparison of dose profiles along the diameter of phantom.

We conducted this experiment to observe how DL-FMP responds to an infeasible (unachievable) dose. We prepared a test case in naïve geometry. The target is a cylinder with 10cm diameter in axial plane and 10cm height along superior-inferior direction, which is in the center of 30cm-diameter water-equivalent phantom. Photon fluences irradiated the target from 64 equally-spaced angles as a surrogate of VMAT arc. The fluence maps from all angles have the same uniform intensity in a 10cm×10cm field size covering the beam-eye-view of cylindrical target. The corresponding dose is the feasible (achievable) dose for this naive geometry with rotational symmetry. The dose distribution in central slice is shown in figure 6.a and the profile along diameter is plotted by the red curve in figure 6.c. The achievable dose has a uniform target dose. The speed of dose falloff outside, reflecting the crossfire effect of beams from every direction, is determined by the target size. Any dose with faster falloff speed would be unachievable in this geometry if the target coverage were to be maintained. We obtained the infeasible (unachievable) dose distribution by painting the dose distribution outside target with a faster falloff speed while keeping target dose to be the same (Ahmed et al., 2017). Practically, we used the dose falloff speed outside a 2cm-diameter cylindrical target to extrapolate the dose outside of our 10cm-diameter target. The resulting unachievable dose distribution is shown in figure 6.b and its falloff speed is compared in figure 6.c. Because of energy conservation, no photon fluence can achieve the dose we painted. We will input the projections of two doses into DL-FMP respectively, and observe the output fluence maps.

# 3 Results

## 3.1 Generalizability to multiple sites

Projections of doses of the lung case, liver case and esophagus case in phantom geometry were input into DL-FMP. We calculated the doses by the fluences predicted by DL-FMP in patient geometry and compared them to GT doses (upper red arrow in figure 1). The results of lung, liver and esophagus cases are tabulated in table 1 and table 2, while the results of prostate testing cases and HN training cases were reported in (Ma et al., 2020) and repeated here. Table 1 reveals that the DL-FMP trained by HN plans can generalize to lung, liver, esophagus and prostate with similar GPR values. Table 2 compares the predicted doses to GT doses in each ROI for lung, liver and esophagus cases, which demonstrates the generalization in terms of clinical dosimetric parameters (Ritter et al., 2011).

Table 1. GPRs on multiple sites. The results of 102 HN test cases and 14 prostate test cases were reported in (Ma et al., 2020). Regions where dose is lower than 10% of prescription dose are excluded.

| Site | 2% / 2mm | 3% / 3mm | 5% / 5mm |
|---|---|---|---|
| 102 HN cases | 93.37±5.86% | 98.06±2.64% | 99.75±0.61% |
| 14 Prostate cases | 95.70±4.27% | 99.06±1.75% | 99.91±0.27% |
| Lung case | 98.79% | 99.96% | 100% |
| Liver case | 99.98% | 100% | 100% |
| Esophagus case | 96.42% | 99.52% | 100% |

Table 2. ROI dose statistics. Coverage is defined as the percentage of PTV voxels with dose higher than prescription dose. Homogeneity is defined as $1 - \frac{D_{5\%} - D_{95\%}}{D_{50\%}}$.

| ROI dose parameter | GT dose | Predicted dose |
|---|---|---|
| Lung | | |
| PTV60Gy coverage | 99.9% | 99.8% |
| PTV60Gy homogeneity | 85.5% | 83.9% |
| Esophagus $D_{5cc}$ | 57.3Gy | 58.3Gy |
| Heart $D_{15cc}$ | 53.9Gy | 53.7Gy |
| Left lung $V_{5Gy}$ | 61.2% | 61.1% |
| Right lung $V_{5Gy}$ | 56.5% | 56.8% |
| SpinalCord $D_{15cc}$ | 27.4Gy | 27.5Gy |
| Liver | | |
| PTV54Gy coverage | 98.4% | 92.9% |
| PTV54Gy homogeneity | 97.1% | 94.6% |
| Liver $D_{700cc}$ | 8.3Gy | 8.2Gy |
| Ribs $D_{5cc}$ | 26.7Gy | 26.3Gy |
| Esophagus | | |
| PTV50Gy coverage | 99.9% | 99.5% |
| PTV50Gy homogeneity | 90.7% | 90.8% |
| PTV45Gy coverage | 99.9% | 99.7% |
| PTV45Gy homogeneity | 94.9% | 94.3% |
| Heart $D_{15cc}$ | 40.7Gy | 40.0Gy |
| Stomach $D_{5cc}$ | 45.4Gy | 45.3Gy |
| SmallBowel $D_{120cc}$ | 23.2Gy | 23.0Gy |
| SpinalCord $D_{15cc}$ | 25.0Gy | 25.2Gy |
| Liver $D_{700cc}$ | 14.9Gy | 14.9Gy |

### 3.2 Generalizability to different delivery modalities

Projections of doses of HN partial-arc plan, pelvis partial-arc plan and HN IMRT plan were input into DL-FMP. The outputs (predicted fluence maps) are compared with GT fluence maps in figure 7, figure 8 and figure 9. The subfigure (a) in the three figures shows the projections of doses as a function of angles. Because each projection is 2D image, we only show the 1D projection of dose in the central axial slice along Y-axis of subfigure (a). The summation of 2D dose projection as a function of angle are shown subfigure (b). The projections from opposite angles are not equal because the projections are weighted by percentage depth dose. The GT and predicted fluence maps are shown in subfigure (c) and (e) respectively. The central row of fluence map, that is the intensity modulation created by the central pair of leaf, is presented in the Y-axis of (c) and (e), corresponding to (a) spatially. The difference of predicted and GT fluence is plotted in subfigure (f). We defined energy fluence as the summation of 2D fluence map from each angle. The angular distributions of energy fluence of GT and predicted fluence maps are compared in subfigure (d).

For the two partial-arc test cases, we focus on if the DL-FMP can identify the beam angles not covered by the arc by outputting empty fluence maps (zero intensity) in those angles. The subfigures (c,d,e,f) of figure 7 and figure 8 demonstrate that the DL-FMP trained by full-arc plans can generalize to partial-arc plans by outputting empty fluence maps for the angles not covered by partial arcs. To evaluate the prediction accuracy dosimetrically, the doses of predicted fluence map (predicted doses) were compared with GT doses. The GPRs of predicted doses are presented in table 3. GPRs of the two partial-arc plans are all higher than 90%, showing excellent generalization. The pelvis case has higher GPR values than HN case because its target size is much smaller (16cc vs 750cc), in which scenario it's easier to pass.

For the IMRT case, we focus on if the DL-FMP can identify the eight beam angles used by the IMRT plan. As shown by the subfigure (c,d,e,f) of figure 9, the predicted fluence is difference from GT fluence. The GT fluence is sparsely distributed in eight angles while the predicted fluence has a broad distribution from all directions in a VMAT fashion. However, as shown by figure 9.d, the energy fluence concentrates at the eight beam angles used by IMRT plan, which indicates that the DL-FMP can identify the eight angles though the output is in the style of VMAT. Even though the predicted and GT fluence are not similar, their doses are surprisingly similar. As shown in the bottom row of table 3, the GPR is higher than 90% with 5mm/5% criterion. A further analysis of the generalizability to IMRT is in Section 4 Discussion.

The results on various delivery modalities show that the DL-FMP, although trained only by full-arc plans, can generalize to partial-arc plan in terms of predicting accurate partial-arc fluence maps and doses. The DL-FMP cannot generalize to IMRT plans in that the predicted fluence is still angularly continuous instead of being discrete like IMRT. However, the predicted dose can be close to GT IMRT dose if evaluated by a large tolerance (5mm/5%).

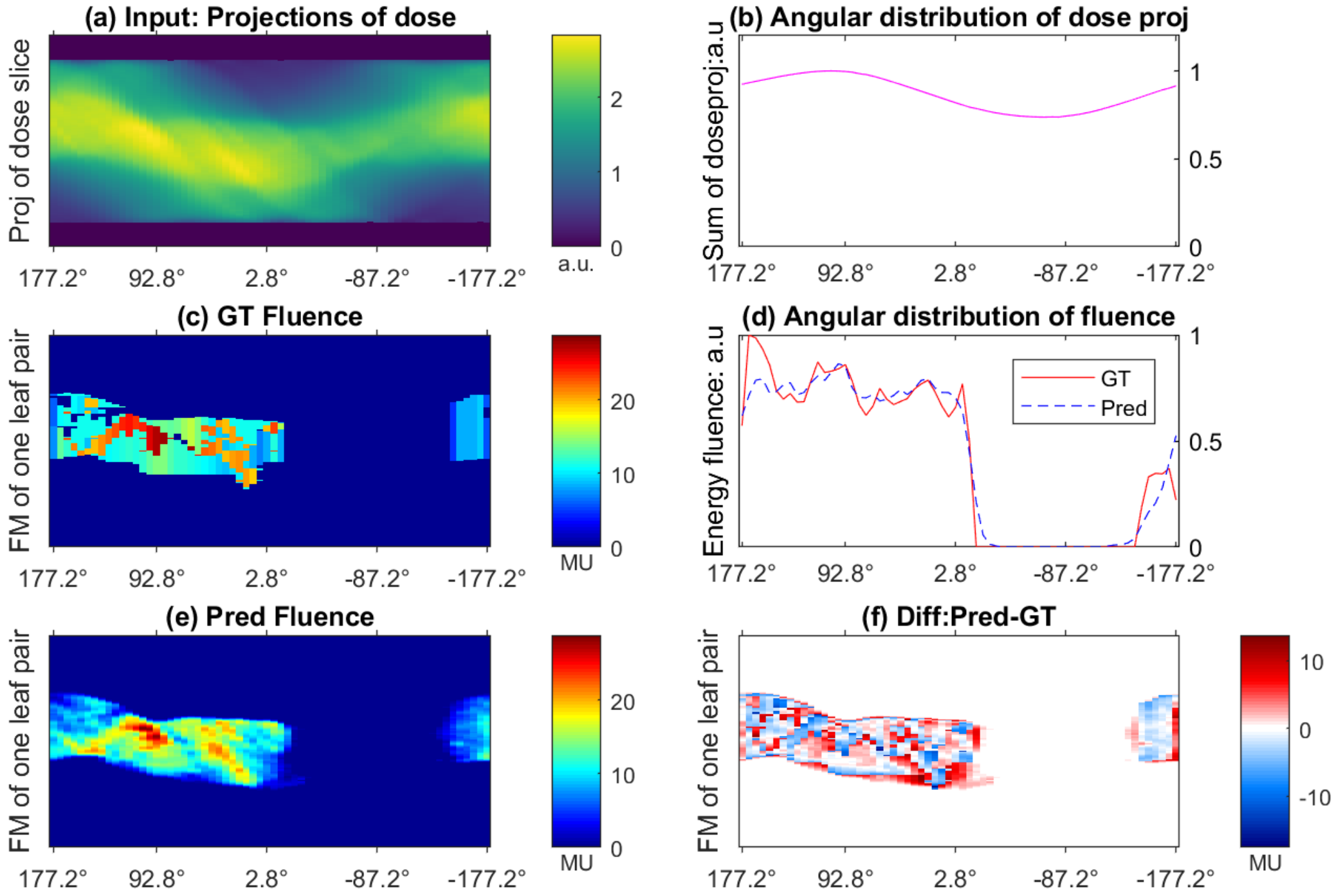

Figure 7. Results of HN partial-arc VMAT plan. (a) Projection of dose in the central slice (y-axis) as a function of projection angle (x-axis). Position of central slice is determined by the position of central leaf on iso-center plane. (b) Summation of pixel values of dose projection over the 2D projection plane (y-axis) as a function of projection angle (x-axis). (c) GT fluence of central leaf pair (y-axis) as a function of beam angle (x-axis). (d) Summation of bixel values of 2D fluence map (y-axis) as a function of beam angle. (e) Predicted fluence of central leaf pair (y-axis) as a function of beam angle. (f) The difference between prediction fluence and GT fluence (subfigure (e) subtracted by (c)).

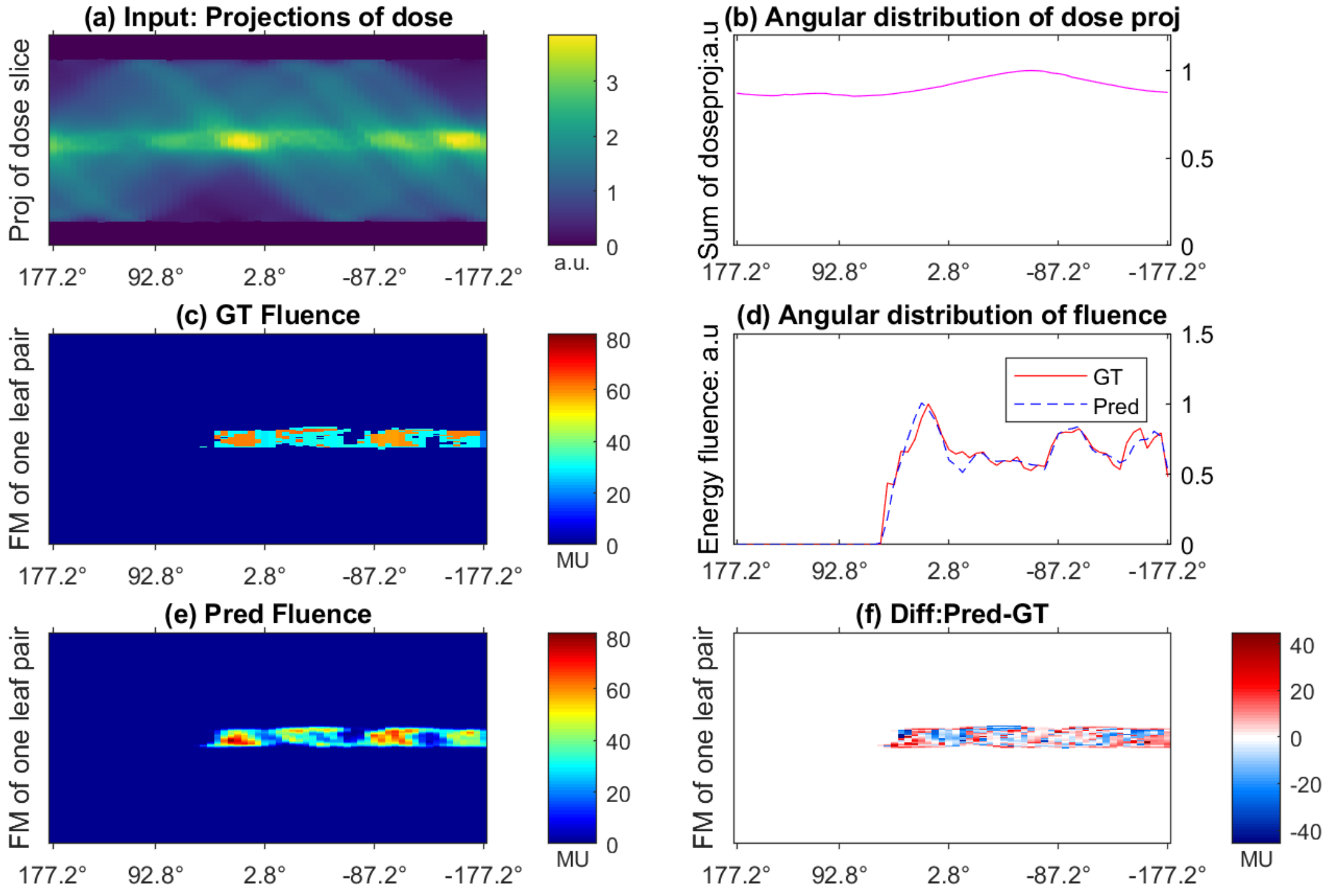

Figure 8. Results of pelvis partial-arc plan. The explanation of each subfigure is same as figure 7.

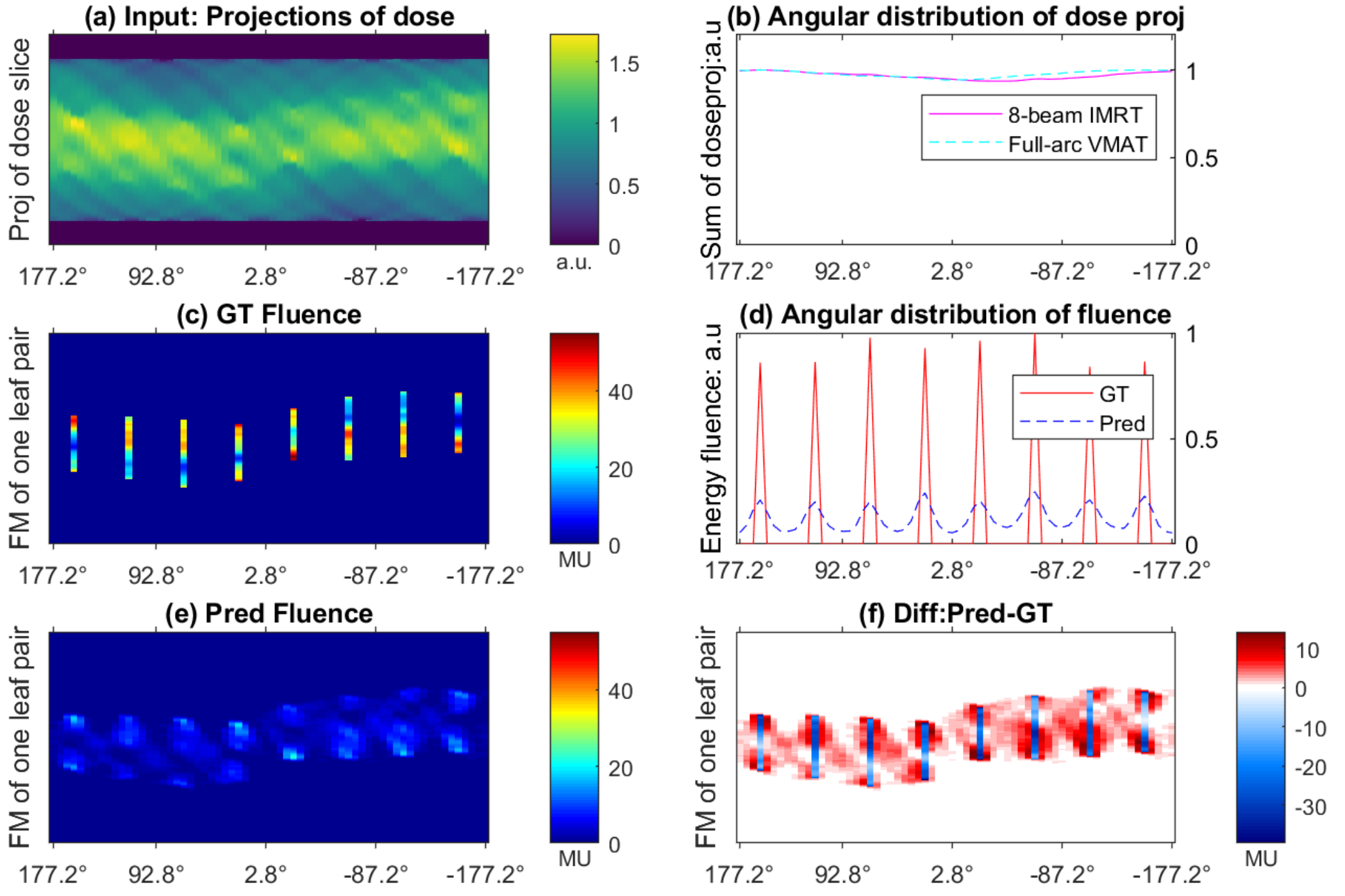

Figure 9. Results of HN 8-beam IMRT plan. . The explanation of each subfigure is same as figure 7. The angular distribution of original clinical VMAT plan is also plotted in (b) for comparison.

Table 3. GPR of predicted doses for plans of different delivery modalities. Regions where dose is lower than 10% of prescription dose are excluded.

| Case | 2% / 2mm | 3% / 3mm | 5% / 5mm |
|---|---|---|---|
| HN partial-arc | 91.05% | 96.90% | 99.58% |
| Pelvis partial-arc | 99.13% | 99.99% | 100.00% |
| HN 8-beam IMRT | 68.86% | 82.73% | 93.11% |

### 3.3 Generalizability to various DOM

The four plans with various DOM in Section 2.3.3 were input to DL-FMP. Predicted and GT fluence maps are compared in figure 11. The differences of four GT fluence maps in DOM is clearly visualized in the first row of figure 11. We also found that the predicted fluence maps is similar to GT correspondences not only in image level, but also in DOM. We can observed that the fluences predicted for plans with less DOM are smoother. To quantitatively measure the DOM, we calculated the summed positive gradient (SPG) of GT and predicted fluences (Craft et al., 2007). The leaf angle is zero degree so the SPG is calculated along X axis of field. Their values are presented in the first row of table 4. We can observe the tendency that DL-FMP predicts fluences with less SPG if the input plan has lower DOM.

Table 4. ROI dose statistics for the GT doses and predicted doses for four DOMs. Coverage is defined as the percentage of PTV voxels with dose higher than prescription dose. Homogeneity is defined as $1-\frac{D_{5\%}-D_{95\%}}{D_{50\%}}$. Mean dose is reported for each OAR in this table.

| | High DOM | | Medium DOM | | Low DOM | | Very low DOM | |
|---|---|---|---|---|---|---|---|---|
| | GT | Pred | GT | Pred | GT | Pred | GT | Pred |
| SPG (a.u) | 114 | 68 | 69 | 59 | 45 | 47 | 32 | 37 |
| PTV70Gy coverage:% | 100.0 | 99.5 | 100.0 | 99.7 | 100 | 99.8 | 99.8 | 99.3 |
| PTV70Gy homogeneity:% | 94.5 | 92.7 | 94.4 | 93.3 | 94.8 | 92.3 | 93.5 | 91.7 |
| PTV59.4Gy coverage:% | 97.9 | 96.2 | 97.6 | 96.0 | 97.3 | 96.0 | 97.6 | 97.1 |
| PTV59.4Gy homogeneity:% | 85.0 | 82.1 | 83.1 | 80.8 | 81.4 | 79.5 | 81.7 | 80.5 |
| PTV56Gy coverage:% | 99.1 | 97.1 | 98.5 | 96.9 | 97.4 | 96.1 | 96.5 | 95.6 |
| PTV56Gy homogeneity:% | 91.0 | 86.8 | 89.1 | 85.3 | 86.3 | 83.4 | 84.4 | 81.9 |
| Left parotid: Gy | 33.5 | 33.1 | 33.4 | 32.9 | 34.6 | 33.7 | 38.0 | 38.2 |
| Right parotid: Gy | 42.5 | 42.4 | 42.8 | 42.6 | 44.5 | 44.6 | 48.3 | 49.2 |
| Spinal cord: Gy | 10.8 | 11.7 | 12.5 | 12.5 | 20.1 | 19.7 | 34.7 | 34.8 |
| Oral cavity: Gy | 19.8 | 20.0 | 20.6 | 20.6 | 21.4 | 21.5 | 22.5 | 22.4 |
| Mandible: Gy | 35.0 | 35.5 | 36.3 | 36.2 | 38.6 | 38.4 | 42.2 | 42.6 |
| Esophagus: Gy | 38.4 | 39.6 | 37.6 | 38.2 | 39.9 | 39.2 | 45.0 | 43.8 |

To study the dosimetric accuracy of predicted fluence maps, we compared the predicted dose (calculated from predicted fluences) and GT dose in patient geometry. The GPRs are tabulated in table 5. The prediction accuracy is slightly better on plans with less DOM and it is always higher than 90% with 3%/3mm criterion for all DOM. DVH curves of GT and predicted doses are compared in figure 10 and ROI dose statistics are compared in table 4.

Table 5. GPRs of predicted dose for various DOMs. Regions where dose is lower than 10% of prescription dose are excluded.

| Case | 2% / 2mm | 3% / 3mm | 5% / 5mm |
|---|---|---|---|
| High DOM | 81.28% | 94.10% | 99.50% |
| Medium DOM | 85.73% | 96.36% | 99.83% |
| Low DOM | 89.04% | 97.72% | 99.99% |
| Very low DOM | 90.72% | 98.20% | 99.99% |

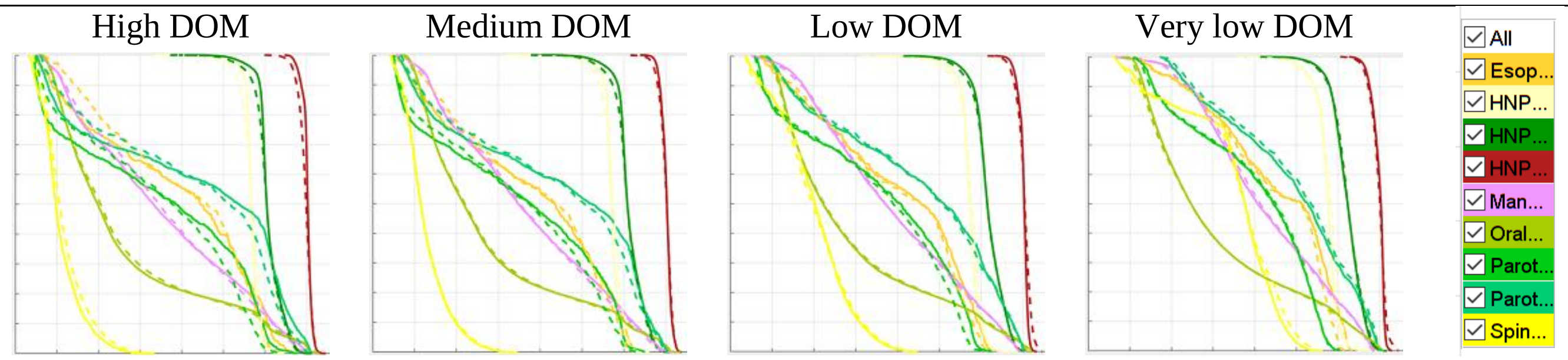

Figure 10. DVHs of the four plans with various DOMs. GT doses are plotted by solid curves and predicted dose are plotted by dashed curves. The color scheme is the same as figure 5.b. The solid curves in four subfigures are the same as the four sets of curves in figure 5.b.

The dosimetric evaluation demonstrated that the DL-FMP trained by clinical plans can generalize to plans in a wide range of DOM. The plans in extreme DOM are beyond the training data with clinically-accepted DOM, as explained in Section 2.3.3. The predicted doses are similar to input doses by GPR, DVH curve and ROI dose statistics. We then go back to check the DOM of predicted fluences and GT fluences. The high-DOM plan was optimized by a penalty on DOM so small that the unnecessary modulation was failed to be suppressed, as shown by the red curve in figure 5.a, dotted pattern in figure 11 (upper left corner) and the large SPG value (114) in table 4. In contrast, the fluences predicted by DL-FMP for high-DOM plan can reproduce the highly-optimal dose while significantly reduce DOM. The SPG of predicted fluence is only 68, decreased by 40% of GT fluence, while the GPR of corresponding doses is as high as 94%. Therefore, the DL-FMP can intelligently adjust the DOM of predicted fluences according to the optimality of input dose, and avoid creating unnecessary modulation in its prediction.

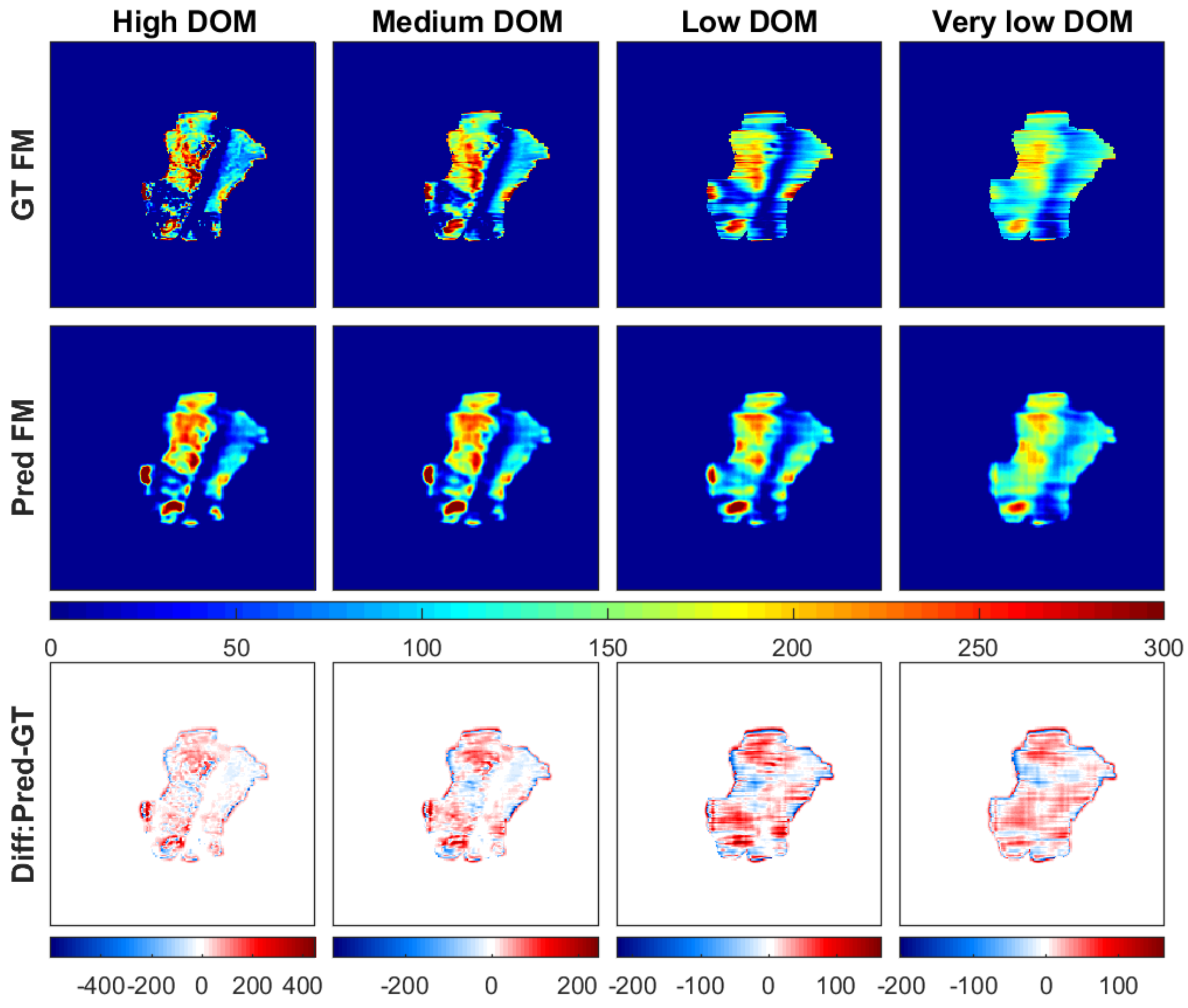

Figure 11. Comparison of GT fluence maps and predicted fluence maps in four DOMs. The unit of all subfigures is MU. For a clear visualization, the GT fluences (first row) and predicted fluences (second row) are displayed in the same color range from 0 to 300MU, although some fluence pixels are higher than 300MU.

### 3.4 Generalizability to infeasible dose

We tested the DL-FMP with the infeasible (unachievable) dose in naïve geometry. The feasible (achievable) dose in the same geometry was also tested for comparison. The inputs, predictions and differences are plotted in figure 12. The difference between achievable dose and unachievable dose are captured by the model input, projections of dose. Their difference are plotted in figure 12.f. The projections of unachievable dose has lower intensity than that of achievable dose by ~10% (0.08/0.8) in the region of target BEV and by ~33% (0.2/0.6) outside of target BEV. Although the input projections of dose are abnormally low, the DL-FMP didn’t crash and still made stable fluence map predictions. Predicted fluence map is displayed in figure 12.g. DL-FMP successfully identified the correct field size and field shape, with near uniform intensity in it. The prediction for achievable dose distribution is shown in figure 12.c and corresponding GT fluence is in figure 12.b. We should note that there’s no corresponding GT fluence map for unachievable dose input. However, we found that DL-FMP can respond to unachievable dose input with fluence maps that can deliver the achievable correspondence. In other words, we input an unachievable dose to DL-FMP, it responded as if it were fed by an achievable dose that corresponds to the unachievable input. The only difference is that the prediction accuracy is worse when the input is unachievable. As shown in figure 12.d&h, the positional accuracy at field boundary and intensity accuracy inside field are worse for unachievable input.

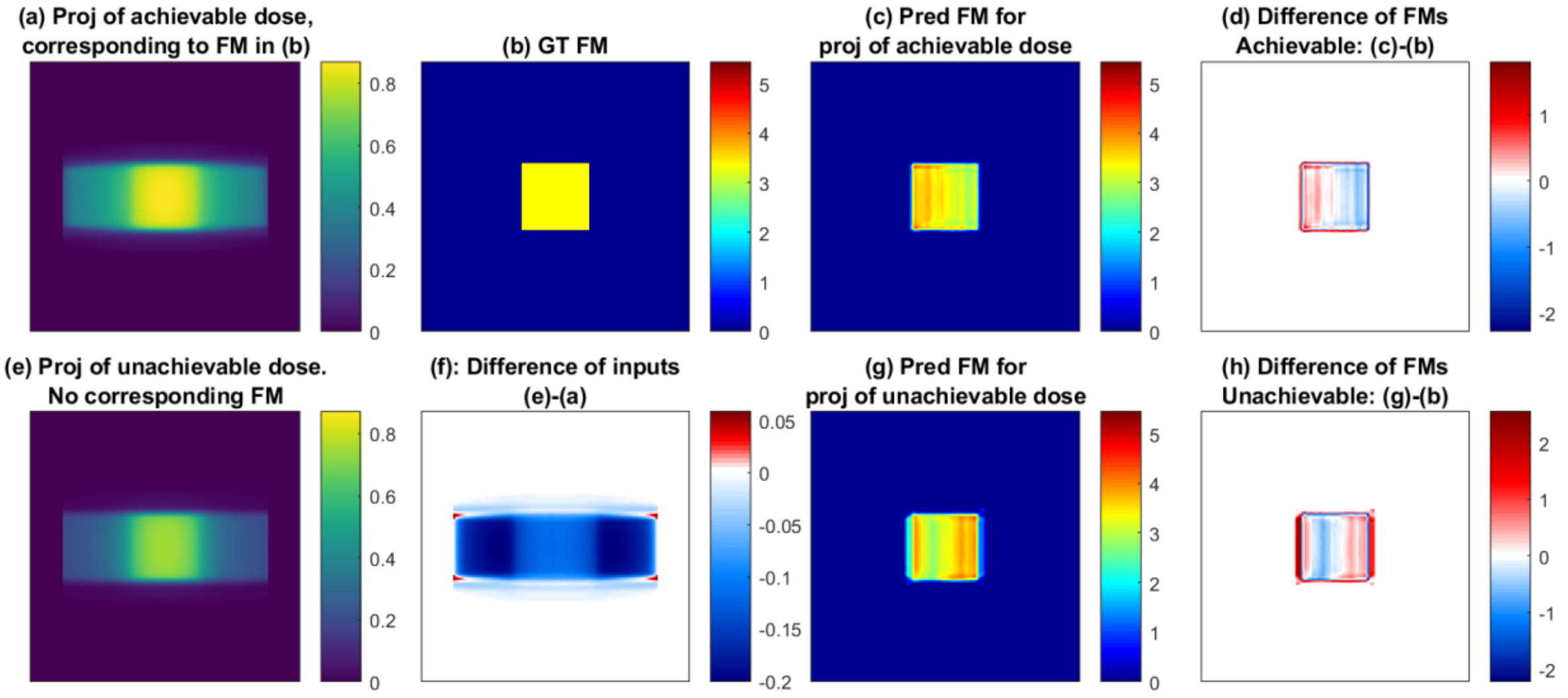

Figure 12. Inputs, GTs, predictions and differences of feasible (achievable) dose and infeasible (unachievable) dose. (a) Projection of feasible (achievable) dose from one angle. (e) Projection of infeasible (unachievable) dose from the same angle. (f) The difference between projections of unachievable and achievable dose. (b) GT fluence corresponding to the achievable dose. (c) Fluence predicted for the achievable dose input in subfigure (a). (g) Fluence predicted for the unachievable dose input in subfigure (e). (d) The difference between predicted and GT fluence for achievable dose. (h) The difference between predicted fluence for unachievable dose and the GT fluence for achievable dose.

We then calculated the dose delivered by predicted fluences. The profiles along diameter of phantom are compared for GT and predicted doses (figure 13). When input is achievable, the predicted dose (green) is close to GT dose (red). While if we input an unachievable dose (blue) with unrealizable dose falloff speed, DL-FMP still managed to output fluences that delivers a dose (magenta) capturing the main features of input unachievable dose, the position and dose level of high dose region covering PTV, although the accuracy is not as high as when input is achievable.

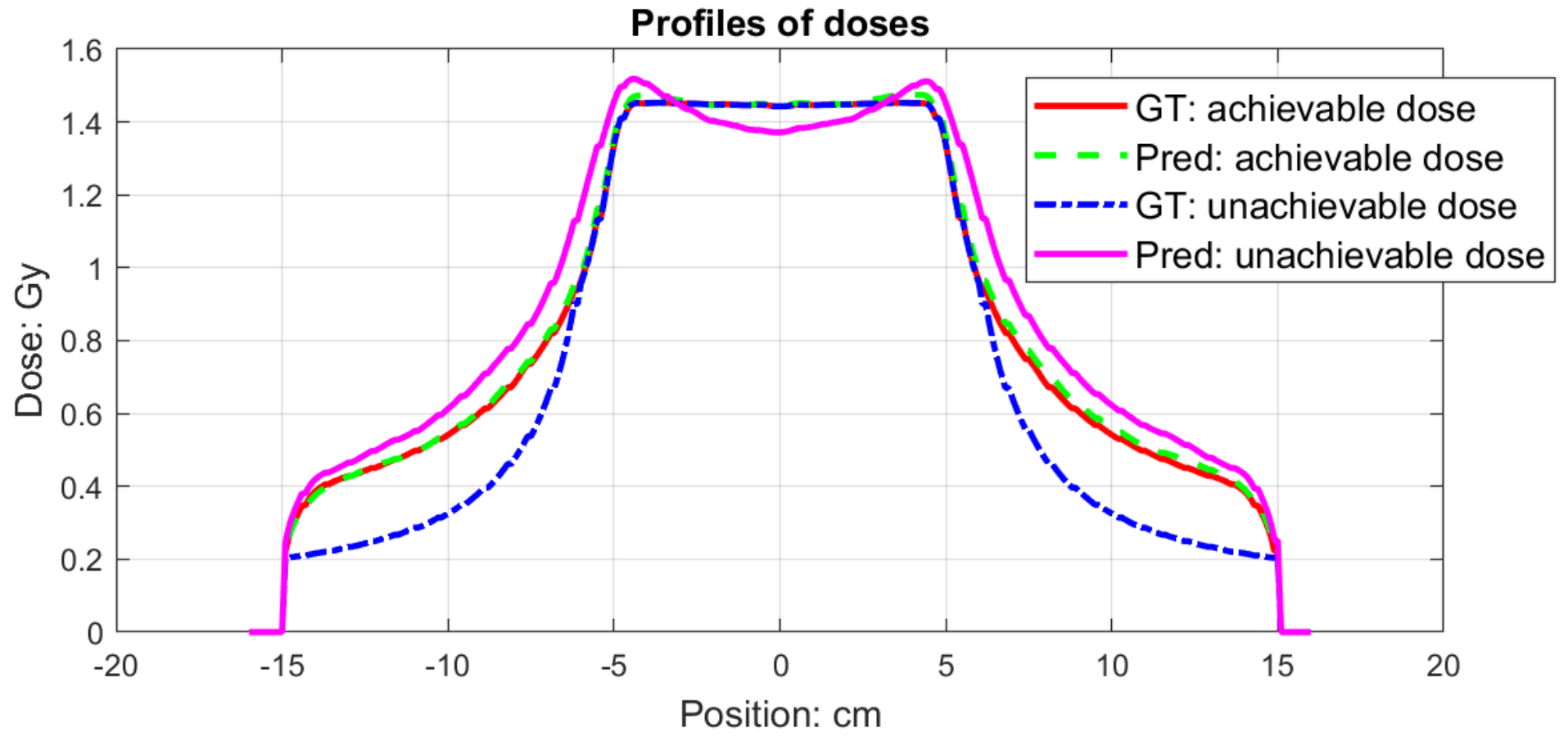

Figure 13. Comparison of dose profiles. The profiles of GT / predicted doses in feasible (achievable) / infeasible (unachievable) cases along the diameter of phantom are compared.

## 4 Discussion

This paper presented a comprehensive experimental validation of the generalizability of DL-FMP. The theory of DL-FMP is based on an inverse mapping $f_P$ from dose projections to fluences, which satisfies $f_P \circ (\widetilde{D}^T D)_{ph} = I$. The theory suggests that the inverse mapping is independent of the specific types of plans, so DL-FMP should be able to generalize to various scenarios not confining to trained data. The DL-FMP was trained by clinical HN full-arc VMAT plans (Ma et al., 2020). This paper manifested the generalizability of the trained DL-FMP to plans for multiple anatomical sites (lung, liver, esophagus and prostate), plans of different delivery modalities (partial-arc VMAT and IMRT) and plans with various DOM. The stability on infeasible (unachievable) dose was also demonstrated by a toy case. These findings established that the DL-FMP is a general inverse planning approach. Similar research on mapping dose to fluence have been conducted for prostate and pancreas (Lee et al., 2019, Wang et al., 2020). However, they only mapped the dose of one beam in an IMRT plan to the fluence map of corresponding beam by a DNN. Actually, it can be achieved by the deconvolution of 2D dose at isocenter plane with a 2D scatter kernel, according to FCBB dose calculation theory (Lu and Chen, 2010). In comparison, DL-FMP is a more general framework for the prediction of a whole VMAT plan with the desired total dose as input.

All the plans in this study are calculated by the commissioning data of a 6MV beam, which was used to generate the training data for the DNN component of DL-FMP. This specific set of beam parameters corresponds to the $(\widetilde{D}^T D)_{ph}$ whose inverse $f_P$ was learned by the DNN. Just as working in patient geometry caused the inconsistency in $D$, different beam parameters (commissioning data and energy) will also introduce a change of $D$ and make the mapping from $P$ to $w$ deviates from the $f_P$ learned by DNN. We applied a fluence scaling method (Ma et al., 2020) as the second component of DL-FMP to address the variation of $D$. Scaling the intensity of fluence can compensate for the variation of $D$ caused by the geometry of transmission medium (electron density distribution), as well as beam properties such as commissioning data and energy. Therefore, generalizability of DL-FMP to different energies and machines is taken care by fluence scaling, which is not included in this paper. We will elaborate the fluence scaling scheme and discuss generalizability in a future paper.

By the experiment on the generalizability to new delivery modalities, we demonstrated that the pre-trained DL-FMP can predict fluence maps for partial-arc VMAT plans directly without re-training. Partial-arc trajectory is often used for tumor located at the single side of body to avoid unnecessary dose to contralateral side. Even though the pre-trained model achieved high accuracy for the two partial-arc test cases, there is space to improve, which might be achieved by including the partial-arc plans into training dataset. The inverse mapping $f_P$ DNN learned still holds in partial-arc plans so there's no confliction between full-arc and partial-arc plans as training data. We have observed an interesting result on IMRT that the DL-FMP took the projections of IMRT dose as input and output fluence maps with angular continuity like VMAT. We shall start a further investigation of IMRT case from examining the input data and looking for the key differences among full-arc VMAT, partial-arc VMAT, and IMRT. The summed intensity of dose projection (input of DL-FMP) as a function of beam angle are plotted in subfigure (b) of figure 7, figure 8 and figure 9. The two partial-arc plans are plotted in figure 7.b and figure 8.b. The IMRT plan was re-optimized from a clinical HN full-arc VMAT case. The re-optimized IMRT and original VMAT are plotted by two separate curves in figure 9.b. Why DL-FMP can generalize to partial-arc plans a technical point of view is elaborated below. Since the projections of dose are weighted by depth, the projections from the angles covered in arc has higher intensity than those not covered since beam entrance dose is higher than beam exit dose and assigned with greater weighting. This tendency can be observed by juxtaposing the subfigure (b) and (d) in figure 7 and figure 8. The intensity variation of dose projections is about 10~30% as shown in subfigure (b). We designed the DNN such that the projections from opposite angles are input into two channels which are convoluted simultaneously (Ma et al., 2020). The intensity difference of opposing dose projections are thus captured efficiently by DNN, which results in successful prediction for partial-arc plans. For the IMRT plans, we found that the eight beam angles cannot be identified from the angular distribution of dose projections. As shown by figure 9.b, the IMRT and original VMAT plan has the same flat-shaped angular distribution. There's a contrastive difference in fluence maps (discrete/continuous angular distribution) between IMRT and VMAT. In contrast, the dose projections of IMRT and VMAT are quite similar. Actually it is understandable from optimization point of view. They are two sides of the same coin. IMRT is a discrete version of VMAT while VMAT is a continuous version of IMRT. The progressive resolution strategy of VMAT optimization (Otto, 2008) starts from a coarse angular resolution similar to the beam configuration of IMRT. It progressively increases number of beams by distributing the fluence of one beam angle to neighboring angles while keeping the dose similar to what it was in coarse resolution. The optimization of IMRT plan is a reverse process of it, which is demonstrated by the simultaneous optimization of beam angles and fluence maps (Jia et al., 2011). It starts from a VMAT-like plan with 180 beams. In each iteration, the beams whose fluence map intensities are smaller than their neighbors will be discarded and the fluence maps of the remaining beams will be finely tuned to maintain the quality of dose as it was in fine resolution. Therefore, the radical difference between IMRT and VMAT is in the discretization/continuity of beam delivery, while the dose distribution of them could be similar in the beam-intersection region. In the eyes of DL-FMP, it's very difficult to tell IMRT and full-arc VMAT by the input projections of dose (figure 9.b). Moreover, all the training data it has seen were VMAT plans. So it is explainable that the DL-FMP predicted fluence maps with the style of VMAT for the test IMRT case. Though it cannot predict fluences in discrete angles like IMRT, its prediction is still a viable solution to the input IMRT dose. The predicted dose (calculated from the predicted fluences in VMAT style) is compared with input IMRT dose in the figure 14.a. Since the PTVs and OARs are mostly in the cross-fire region, the two doses inside ROIs are close and so as their DVH curves. The gamma map is presented in figure 14.b, which clearly illustrates that the disagreement (voxels with gamma value above one) between predicted dose (from VMAT-style fluence prediction) and GT IMRT dose is mainly distributed in beam-entrance region.

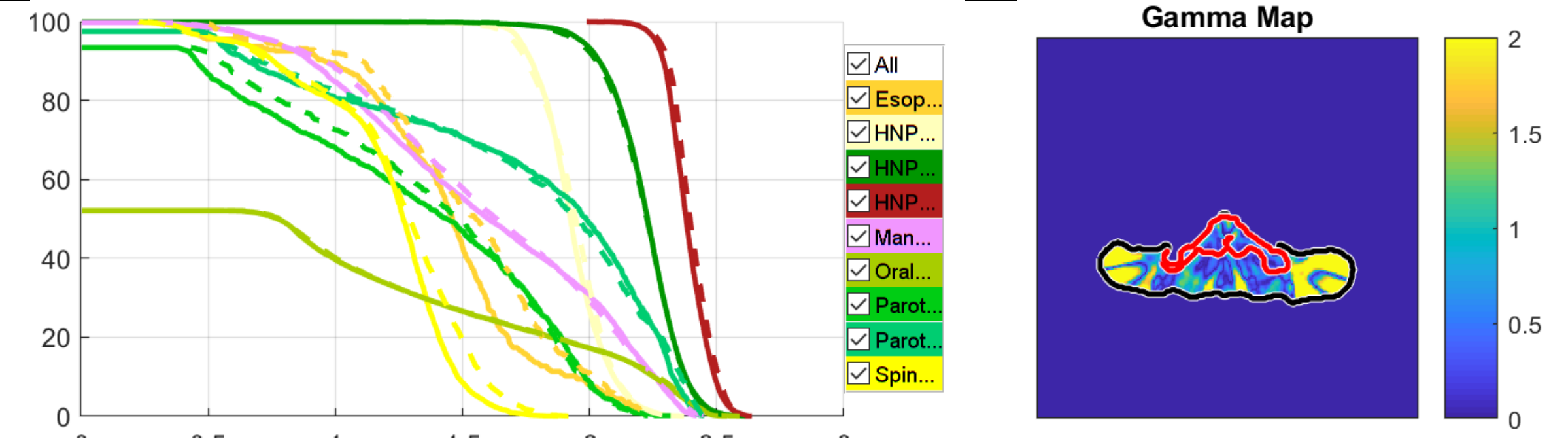

(a) DVH of predicted dose and GT dose (b) Gamma map

Figure 14. Prediction of DL-FMP on 8-beam IMRT plan. (a) DVH curves of GT IMRT dose (solid) and predicted dose (dashed). (b) Gamma map (3%/3mm). PTV is in red contour. Voxels with gamma values lower than one pass. We used a color range of [0,2] for better visualization, although gamma value could be higher than two.

The fluence maps of coplanar VMAT plans are discretized as 64 equi-angular coplanar control points, which required 10GB GPU memory for model training (Ma et al., 2020). The angular resolution can be further increased by using GPU with larger memory or by designing memory-efficient model structure. For the time being, non-coplanar plans are not under investigation since the representation of non-coplanar trajectory is not supported by DL-FMP. We will further investigate if DL-FMP can accommodate non-coplanar plans.

Conventional planning workflow could be inefficient for complicated patient cases with many conflicting objectives. To reduce the back-and-forth between decision maker (physicians) and plan maker (dosimetrists), pioneer researchers suggested the use of interactive planning (Otto, 2014). This is a process of interactively tuning dose distribution and exploring trade-offs by physicians, which is separated from deriving actual delivery parameters for the physician-desired dose by dosimetrists. Interactive planning is initially realized by fast plan optimization using an approximate dose calculation engine (Otto, 2014). With the advent of deep learning, predicting volumetric dose distributions became possible and researchers focused on how to leverage the flexibility of DNN to generate a set of dose distributions for physicians to choose from (Nguyen et al., 2019a, Nilsson et al., 2021, Ma et al., 2019). Although predicting doses as such may be faster than obtaining plans by optimization, the drawback is that we cannot guarantee if the predicted doses are achievable and optimal. Additional dose mimicking is usually adopted to resolve this issue (McIntosh et al., 2017, Nilsson et al., 2021), but at the price of wasting the speed advantage of DNN-based dose prediction. The proposed DL-FMP has the potential to replace dose mimicking optimization. Conceptually, the predicted dose, if in phantom geometry, can be input to DL-FMP, which will predict corresponding fluence maps. The computation time is less than 1 second (Ma et al., 2020), much shorter than optimization. The predicted fluence maps can be used to calculate a fluence-based dose distribution, to ensure that the resulting dose of interactive tuning is feasible. DOM of predicted fluences can indicate the optimality of dose, allowing for the exploration of trade-off between plan complexity and plan quality.

## 5 Conclusion

DL-FMP is an inverse planning approach that inversely map the projections of desired dose to fluence maps. In this work, we experimentally validated the generalizability of the DL-FMP which was trained by clinical HN full-arc VMAT plans. We demonstrated with controlled experiments that DL-FMP can generalize to plans for multiple anatomical sites (lung, liver, esophagus and prostate), to plans in different delivery modalities (partial-arc VMAT and IMRT), and to plans with various DOM. DL-FMP can even predict fluence maps that preserve the main dosimetric features for infeasible input dose. These validated generalizabilities agree with the theory basis of DL-FMP, a general inverse mapping. In conclusion, DL-FMP sets up a general framework that predicts fluence map for desired dose distribution. The generalizability allows for broad clinical implementation.

## 6 Acknowledgements

This work was supported in part by NIH grants (R01 CA235723, R01 CA218402). Jonathan Feinberg edited the manuscript.

[†]Corresponding Author: Xuejun.Gu@UTSouthwestern.edu Weiguo.Lu@UTSouthwestern.edu